\documentclass[aps,prd,letterpaper,showkeys,showpacs,twocolumn,nofootinbib,preprintnumbers,longbibliography,unsortedaddress]{revtex4-1}

\usepackage[top=2.8cm, bottom=2.8cm, left=2.4cm, right=2.4cm]{geometry}
\usepackage[T1]{fontenc} 
\usepackage{amsmath,amssymb,lmodern,amsfonts} 
\usepackage{graphicx}
\usepackage{hyperref}
\usepackage{color}
\usepackage{natbib}
\usepackage[caption=false]{subfig}
\usepackage[export]{adjustbox}
\usepackage{geometry}
\usepackage{booktabs}
\usepackage{amstext}

\usepackage{varwidth}
\usepackage{tikz}

\definecolor{darkgreen}{rgb}{0,0.6,0}



\begin{document}
\title{Reconstructing the parameter space of non-analytical cosmological fixed points}
\author{Santiago Garc\'ia-Serna}
\email{santiago.serna@correounivalle.edu.co}
\author{J. Bayron Orjuela-Quintana}
\email{john.orjuela@correounivalle.edu.co}
\author{C\'esar A. Valenzuela-Toledo}
\email{cesar.valenzuela@correounivalle.edu.co}
\author{Hernán Ocampo Durán}
\email{hernan.ocampo@correounivalle.edu.co}
\affiliation{Departamento de F\'isica, Universidad del Valle, \\ Ciudad Universitaria Mel\'endez, Santiago de Cali 760032, Colombia}

\begin{abstract}
Dynamical system theory is a widely used technique in the analysis of cosmological models. Within this framework, the equations describing the dynamics of a model are recast in terms of dimensionless variables, which evolve according to a set of autonomous first-order differential equations. The fixed points of this autonomous set encode the asymptotic evolution of the model. Usually, these points can be written as analytical expressions for the variables in terms of the parameters of the model, which allows a complete characterization of the corresponding parameter space. However, a thoroughly analytical treatment is impossible in some cases. In this work, we give an example of a dark energy model, a scalar field coupled to a vector field in an anisotropic background, where not all the fixed points can be analytically found. Then, we put forward a general scheme that provides a numerical description of the parameter space. This allows us to find interesting accelerated attractors of the system with no analytical representation. This work may serve as a template for the numerical analysis of highly complicated dynamical systems.
\end{abstract}



\maketitle

\section{Introduction} 
\label{Introduction}

Since the discovery of the accelerated expansion of the Universe at the end of the last century \cite{SupernovaSearchTeam:1998fmf,SupernovaCosmologyProject:1998vns,SupernovaSearchTeam:1998bnz}, a wealth of observations have established the $\Lambda$CDM model, which considers the cosmological constant $\Lambda$ and a Cold Dark Matter (CDM) component, as the simplest and most accurate description of the evolution of the Universe from the Big-Bang to nowadays \cite{Boomerang:2000efg, WMAP:2003elm, SDSS:2003eyi, Jaffe:2003it, SDSS:2006lmn, Percival:2007yw, Aubourg:2014yra, Planck:2018vyg}. However, some pretty interesting discrepancies, such as the $H_0$ and $\sigma_8$ tensions \cite{Perivolaropoulos:2021jda, Riess:2021jrx, Freedman:2021ahq, Abdalla:2022yfr, PhysRevD.91.103508, Gatti:2021uwl, PhysRevLett.111.161301, Zurcher:2021bjz, Blanchard:2021dwr, Huang:2021tvo, Heymans:2020gsg, DES:2022ign} or the Cosmic Microwave Background (CMB) anomalies \cite{Bennett:2010jb, Perivolaropoulos:2014lua, Schwarz:2015cma, Planck:2019evm, Planck:2019kim}, are challenging this paradigm. These tensions have provided further motivation to explore theoretical alternatives to the so-called concordance model \cite{DiValentino:2017iww, Guo:2018ans, diValentino:2021izs, Heisenberg:2022lob}.

One of the most powerful tools used in the analysis of the parameter space of a theoretical proposal is the dynamical system technique \cite{Copeland:2006wr, Wainwright2009, Garcia-Salcedo:2015ora}. In the realm of cosmology, this technique consists of recasting the equations of motion of the fields in terms of dimensionless variables, which are typically chosen from the first Friedman equation. This yields a set of autonomous first-order differential equations whose stationary solutions lie in the phase space of the dynamical variables. This technique has been widely used in cosmology (see e.g., Refs. \cite{Bahamonde:2017ize, Alvarez:2019ues, Guarnizo:2020pkj, Motoa-Manzano:2020mwe, Orjuela-Quintana:2020klr, DeFelice:2016yws, Basilakos:2019dof}). 

In a nutshell, the autonomous set can be schematically represented as
\begin{equation}
\{ x'_i = f_i (x_1, \ldots, x_n; a_1, \ldots, a_k)\, | \ i = 1, \ldots, n\},
\label{Eq: Autonomous System}
\end{equation}
where the prime denotes the derivative with respect to the number of $e$-folds, and $f_i$ is an algebraic expression in terms of the $n$ variables $x_j$ and the $k$ parameters $a_j$ of the model. The possible stationary states of this set, i.e., where $\{x_i' = 0 \ | \ i = 1, \ldots, n\}$, define the so-called fixed points of the system, which can be found by solving the set of algebraic equations given by $\{ f_i (x_1, \ldots, x_n ; a_1, \ldots, a_k) = 0 \ | \ i = 1, \ldots, n \}$. Usually, these algebraic equations are analytically solvable. The solutions correspond to expressions for the variables in terms of the parameters, allowing a full characterization of the parameter space of the model. However, there are some cases where these solutions are not available, forbidding a complete examination of the parameter space. For instance, any of the equations $f_i = 0$ involves a polynomial of degree greater than 4 in the dynamical variables, which implies that analytical solutions do not exist due to the fundamental theorem of Galois theory \cite{ribes2010free}.

Although analytical fixed points are not available in all cases, it is still possible to solve numerically the algebraic set  $\{ f_i(x_1, \ldots, x_n) = 0 \ | \ i = 1, \ldots, n \}$ assuming specific values for the parameters. In this work, we put forward a general framework to determine the stability of these ``numerical fixed points'' given a finite portion of the parameter space. In the particular case of a scalar field coupled to a vector field evolving in an anisotropic background, our numerical scheme allows us to estimate the attraction region in the parameter space of anisotropic accelerated solutions which do not have analytical counterparts. For this model, we numerically solve the corresponding autonomous set assuming some values of the parameters, finding general agreement with the asymptotic behaviors predicted for the numerical fixed points.

This paper is organized as follows. In section \ref{Sec: Dynamical systems: numerical approach}, we provide a general overview of our numerical implementation of dynamical systems. We then proceed to illustrate our scheme by applying it to a particular cosmological model in section \ref{Sec: Illustration of the Method}. In section \ref{Sec: Numerical Integration of the Autonomous Set}, we present a numerical solution for the full autonomous set and compare with the predictions from the numerical fixed points. Finally, our conclusions are summarized in section \ref{Sec: Conclusions}.

\section{Dynamical systems: numerical approach}
\label{Sec: Dynamical systems: numerical approach}

In this section, we describe our general framework to analyze the stability properties of a region in the parameter space of a model with no analytical fixed points. Assuming an autonomous system of $n$ differential equations, as in Eq. \eqref{Eq: Autonomous System}, our numerical scheme relies on the following steps:

\begin{itemize}
\item[1.] Choice of a representative region in the parameter space of the model.
\end{itemize}

This can be done by defining an interval for each parameter, i.e., $a_{i_\text{min}} < a_i < a_{i_\text{max}}$ for $i = 1, \ldots, k$, such that the Cartesian product of these intervals build a portion of the parameter space. The length of each interval should take into consideration relevant physical conditions of the model and the symmetries of the autonomous system. 

\begin{itemize}
\item[2.] Setting of physical constraints to discriminate between cosmologically viable and non-viable fixed points.
\end{itemize}

These constraints are required to ensure that the numerical solutions of the autonomous set are related to some physically interesting properties of the model. For instance, imposing the condition $w_{\text{eff}} < -1/3$, where $w_\text{eff}$ is the effective equation of state, implies that any found solution describes an expanding universe at an accelerated rate. Other conditions have to be satisfied for the consistency of the theory, for instance, the density parameters must take on nonnegative values. Numerical solutions meeting all the chosen conditions are called ``cosmologically viable''. Otherwise, they are cataloged as ``non-viable solutions'' and discarded. Note that all the cosmologically viable solutions coming from the same set of parameters are physically indistinguishable.

\begin{figure}[t!]
\centering
\includegraphics[width = \linewidth]{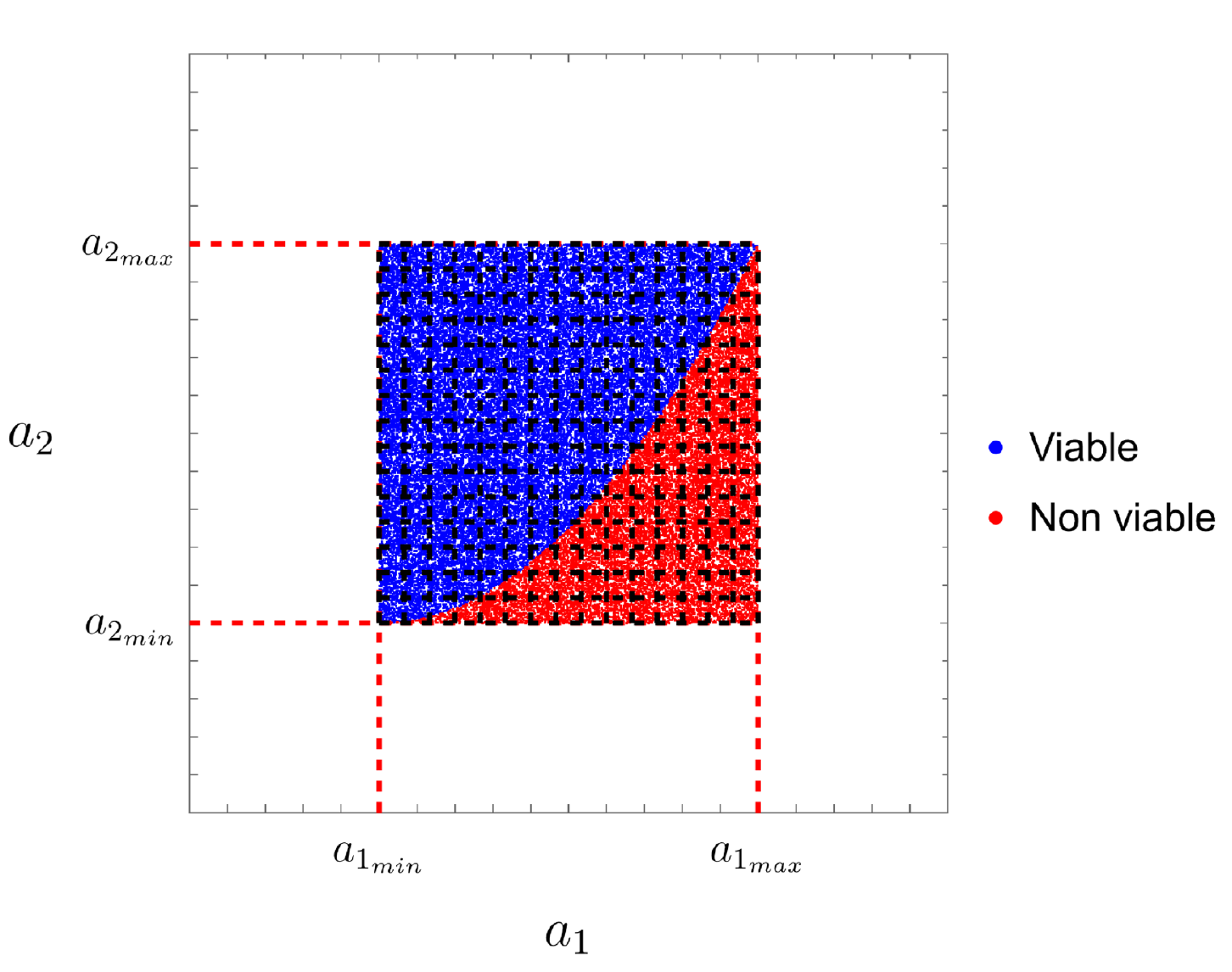}
\caption{Parameter space where a stochastic search for viable and non-viable points has been performed.}
\label{Fig: Stochastic Search}
\end{figure}

\begin{itemize}
\item[3.] Stochastic search of viable solutions in the chosen region in the parameter space.
\end{itemize}
A large number $N$ of random points in the parameter space $\{a_1, \ldots\, a_k\}$ is generated. The corresponding fixed points for each set of parameters are found by numerically solving the system of algebraic equations $\{ f_i(x_1, \ldots, x_n) = 0 \ | \ i = 1, \ldots, n \}$.\footnote{This can be done using the default numerical methods in the \texttt{NSolve} of \texttt{Mathematica}, for example.} Then, the chosen physical constraints in the previous step are evaluated for all the solutions, such as those which fulfill them are cataloged as ``viable solutions'' and the others are discarded. In the case when the model has two parameters, this stochastic search can yield to results schematically similar to the region plotted in FIG.~\ref{Fig: Stochastic Search}. In our case of interest, all the viable points will correspond to accelerated solutions of the system, i.e., Dark Energy (DE) dominated points.
    
\begin{itemize}
\item[4.] Stability analysis of the region of viable points. 
\end{itemize}

In general, the stability of a fixed point can be determined by computing the real part of the eigenvalues of the Jacobian matrix evaluated in the point.\footnote{The Jacobian matrix for a dynamical system with $n$ equations  of the form $x'_i = f_i(x_1, \ldots, x_n)$ is defined as $J_{i j} \equiv \frac{\partial f_i}{\partial x_j}$.} When all the corresponding eigenvalues are negative, we say that the fixed point is an attractor. When all of them are positive, the fixed point is a source or repeller. If there is a mix of negative and positive eigenvalues, then the fixed point is a saddle. 

In the cosmological context, a proper expansion history requires the existence of at least three fixed points as follows: $i)$ a radiation dominated point which can be a saddle or a source, $ii)$ a saddle matter dominated point, and $iii)$ a DE dominated point that can be an attractor of the system. 

Note that for a given point in the parameter space, there could exist several fixed points with their own stability; for example, the three stages of dominance (radiation, matter, and DE epochs) should exist for a given set of parameters in order for the model to be able to reproduce a correct expansion history. Therefore, it is necessary to compute all the available Jacobian matrices (and their eigenvalues) for each set of parameters yielding cosmologically viable solutions. Since viable solutions from a given set of parameters have the same physical characteristics, the stability of the point in the parameter space is determined by the most stable point, that is:

\begin{itemize}
\item If all eigenvalues of at least one of the matrices are negative, then this point is an attractor and the other possible fixed points are discarded.
\item If all eigenvalues of all matrices are positive, then the stability corresponds to a repeller.
\item If the stability is not an attractor neither a repeller in the sense described above, one or more matrices have a mix of positive and negative eigenvalues, and the stability is of a saddle point.
\end{itemize}

Here, we are mainly interested in DE dominated attractor points. Therefore, once an attractor is found, the computation of the Jacobian matrices for the given set of parameters is stopped, just to proceed to another point in the parameter space.

In the following section, we will illustrate the numerical method described here by analyzing the asymptotic behavior of a specific model, prioritizing the search of DE domination points. 

\section{Illustration of the Method}
\label{Sec: Illustration of the Method}

\subsection{Tachyon field in an anisotropic background}

The Lagrangian of our template model is given by 
\begin{equation}
S \equiv \int \text{d}^4 x \sqrt{- g} \left( \mathcal{L}_\text{EH} + \mathcal{L} + \mathcal{L}_m + \mathcal{L}_r \right),
\label{Eq: Tachyon Field Action}
\end{equation}
where $\mathcal{L}_\text{EH} \equiv M_\text{Pl}^2 R/2$, $M_\text{Pl}$ is the reduced Planck mass, $R$ is the Ricci scalar, $\mathcal{L}_m$ and $\mathcal{L}_r$ are the Lagrangians for matter and radiation, respectively, and 
\begin{eqnarray} 
\mathcal{L} \equiv - V(\phi)\sqrt{1+\partial_{\mu}\phi\partial^{\mu}\phi} - \frac{1}{4} f(\phi) F^{\mu\nu} F_{\mu\nu},
\label{Eq: Tachyonic Lagrangian}
\end{eqnarray}
where $\phi$ is the scalar tachyon field, $V(\phi)$ is its potential, $F_{\mu\nu}\equiv \nabla_{\mu}A_{\nu}-\nabla_{\nu}A_{\mu}$ is the strength tensor associated to the vector field $A_{\mu}$, and $f(\phi)$ is a coupling function between $\phi$ and $A_\mu$. Next, we assume that the background geometry is described by a Bianchi I metric with rotational symmetry in the $y-z$ plane, such as the line element is written as 
\begin{equation}
\text{d} s^2 = - \text{d} t^2 + a^2(t)\left[e^{-4\sigma(t)}\mathrm{d} x^2 + e^{2\sigma(t)} \left( \mathrm{d} y^2 + \mathrm{d} z^2 \right)\right],
\label{Eq: Background Metric}
\end{equation}
where $a(t)$ is the average scale factor and $\sigma(t)$ is the geometrical shear, being both functions of the cosmic
time $t$. In order to preserve the symmetries of the background, we choose the field profiles as
\begin{equation}
\phi \equiv \phi(t), \quad A_{\mu} \equiv \left(0, A(t), 0, 0\right),
\label{Eq: Field Profiles}
\end{equation}
being $A(t)$ a scalar field and the unique component of the vector field. 

We want to make a few comments about the specific model we study here. Firstly, dynamical system analysis of theories involving a scalar tachyon field can be found in the literature, Refs. \cite{Aguirregabiria:2004xd, Abramo:2003cp, Bagla:2002yn, Nozari:2013mba, Hussain:2022dhp} are some examples. All these works get analytical results. However, as pointed out in Refs.  \cite{Ohashi:2013pca} and \cite{ Orjuela-Quintana:2021zoe}, a full analytical description of the dynamical system is impossible for some non-canonical scalar field models in a Bianchi-I background, such as the  Dirac-Born-Infeld (DBI) field. Since the Lagrangian of the DBI model shares some similarities  with the Lagrangian in Eq. \eqref{Eq: Tachyonic Lagrangian}, we expect that the dynamical system of this model also lacks of a full analytical description. We will explicitly show that the  anisotropic dark energy attractor of our model cannot be studied analytically and use the numerical framework explained in Sec. \ref{Sec: Dynamical systems: numerical approach} to reconstruct a portion of its parameter space. 

\subsection{Dynamical System} 

Following the standard procedure \cite{Garcia-Salcedo:2015ora}, we derive the evolution equations of the model by varying the action in Eq. \eqref{Eq: Tachyon Field Action} with respect to the metric, the scalar field and the vector field. After these variations, we replace the Bianchi I metric given in Eq. \eqref{Eq: Background Metric}, and the ansatz for the fields in Eq. \eqref{Eq: Field Profiles} obtaining:
\begin{align}
3 M_\text{Pl}^2 H^2  &= \frac{1}{2} f \frac{e^{4\sigma} \dot{A}^2}{a^2} +\frac{ V }{ \displaystyle\sqrt{1-\dot{\phi}^2}} +\rho_m + \rho_r \label{Eq: 1º Friedman}\\
&+ 3 M_{\text{Pl}}^2 \dot{\sigma}^2,\nonumber     
\\
- 2 M_\text{Pl}^2 \dot{H} &= \dot{\phi}^2 \frac{ V }{ \displaystyle\sqrt{1-\dot{\phi}^2} } + \frac{2}{3} f \frac{e^{4\sigma} \dot{A}^2}{a^2} + \frac{4}{3} \rho_r \label{Eq: 2º Friedman} \\ 
&+ \rho_m + 6 M_{\text{Pl}}^2 \dot{\sigma}^2,\nonumber
\\
\ddot{\sigma} + 3H \dot{\sigma} &= \frac{e^{4 \sigma} f \dot{A}^2 }{ 3 a^2 M_{\text{Pl}}^2 },
\label{Eq: sigma} \\
\frac{\ddot{\phi}}{1 - \dot{\phi}^2 } &=  \frac{ \displaystyle\sqrt{1-\dot{\phi}^2} f_{,\phi} }{ 2 V a^2  }  e^{4\sigma} \dot{A}^2 -\frac{V_{,\phi}}{V}-3H\dot{\phi}, \label{Eq: Eq phi} \\
\frac{\ddot{A}}{\dot{A}} &= -\frac{\text{d}}{\text{d} t}\ln{\left(a f e^{4\sigma}\right)}, \label{Eq: Eq A}
\end{align}
where $H$ is the Hubble parameter, $\rho_r$ and $\rho_m$ are the densities of matter and radiation, respectively, and a dot indicates a derivative with respect to $t$. Equations \eqref{Eq: 1º Friedman} and \eqref{Eq: 2º Friedman} correspond to the first and second Friedman equations, Eq. \eqref{Eq: sigma} is the evolution equation for the geometrical shear, and Eqs. \eqref{Eq: Eq phi} and \eqref{Eq: Eq A} are the equations of motion for the scalar and vector fields, respectively. The above equations can be recast in terms of the following dimensionless variables
\begin{gather}
\notag
x \equiv \dot{\phi}\text{,\hspace{0.4cm}}y^2 \equiv \frac{V(\phi)}{3M_{\text{Pl}}^2H^2}\text{,\hspace{0.4cm}}z^2 \equiv \frac{1}{2}f(\phi)\frac{e^{4\sigma}\dot{A}^2}{3M_{\text{Pl}}^2H^2a^2}, \\
\Omega_m \equiv \frac{\rho_m}{3M_{\text{Pl}}^2H^2}\text{,\hspace{0.4cm}}\Omega_r \equiv \frac{\rho_r}{3M_{\text{Pl}}^2H^2}\text{,\hspace{0.4cm}} \Sigma \equiv \frac{\dot{\sigma}}{H}.
\label{Eq: Dynamical Variables}
\end{gather}
The first Friedman equation in Eq. \eqref{Eq: 1º Friedman} becomes the constraint
\begin{equation}
1 = \frac{y^2}{\sqrt{1 - x^2}} + z^2 + \Sigma^2 + \Omega_m + \Omega_r,
\label{Eq: Friedman Constraint}
\end{equation}
while from the second Friedman equation in Eq. \eqref{Eq: 2º Friedman} we can compute the deceleration parameter $q \equiv - 1 - \dot{H}/H^2$ obtaining
\begin{equation}
 q = \frac{1}{2} \left[1 - 3 \sqrt{1 - x^2} y^2 + z^2 + \Omega_r + 3 \Sigma^2 \right].\label{Eq: q deceleration parameter}
\end{equation}
We can easily integrate Eq. \eqref{Eq: Eq A} such that for the vector field degree of freedom we have 
\begin{equation}
    \dot{A} = c \frac{e^{-4 \sigma}}{a f},
\end{equation}
where $c$ is a constant.
From the Friedman equations \eqref{Eq: 1º Friedman} and \eqref{Eq: 2º Friedman}, we can identify the density and pressure of dark energy, which can be written in terms of the dimensionless variables as
\begin{align}
\rho_{\text{DE}} \equiv& 3M_\text{Pl}^2 H^2 \left( \frac{y^2}{\sqrt{1 - x^2}} + z^2 + \Sigma^2 \right), \\
p_{\text{DE}} \equiv& 3M_{\text{Pl}}^2H^2\left(-y^2\sqrt{1-x^2}+\frac{1}{3}z^2+\Sigma^2\right),
\end{align}
respectively. Now, the equation of state of DE, $w_{\text{DE}} \equiv  p_{\text{DE}} / \rho_{\text{DE}}$, is given by
\begin{equation}
w_{\text{DE}} = -1 + \frac{2}{3} \frac{\frac{3}{2} \frac{x^2 y^2}{\sqrt{1 - x^2}} + 2 z^2 + 3 \Sigma^2}{\frac{y^2}{\sqrt{1 - x^2}} + z^2 + \Sigma^2}.
\end{equation}
Note that we included the geometrical shear $\Sigma$ in the definition of $\rho_{\text{DE}}$ and $p_{\text{DE}}$. This choice allows us to write the DE continuity equation as 
\begin{equation}
\dot{\rho}_{\text{DE}} + 3 H (\rho_{\text{DE}} + p_{\text{DE}}) = 0,
\end{equation}
which is the usual form of the continuity equation of an uncoupled fluid, stressing that we are considering that DE can be an anisotropic fluid.

In order to calculate the evolution equations for the dimensionless variables, we differentiate each variable in Eq. \eqref{Eq: Dynamical Variables} with respect to the number of $e$-folds.\footnote{The relation between the number of $e$-folds and the scale factor is $N \equiv \ln a$.} We get
\begin{align}
 x' &= \sqrt{3} \left(1 - x^{2} \right) \left[\frac{\sqrt{1 - x^{2}}}{y} \beta z^{2} - \sqrt{3}  x - \alpha y \right], \label{Eq: x prime} \\
 y' &= y \left[\frac{\sqrt{3}}{2} \alpha y x + q + 1 \right], \\
 z' &= z (q - 1) - 2 z \Sigma - \frac{\sqrt{3}}{2} \beta x y z, \label{zprime} \\
 \Sigma' &= \Sigma(q - 2) + 2 z^{2}, \label{Eq: Sigma prime}\\
 \Omega_r' &= 2 \Omega_r (q - 1),\label{Eq: Omega prime}
\end{align}
where
\begin{equation}
\alpha(t) \equiv M_\text{Pl} \frac{V_{, \phi}}{V^{3/2}}, \quad 
\beta(t) \equiv M_\text{Pl} \frac{f_{,\phi}}{f \sqrt{V}}.
\label{Eq: alpha and beta}
\end{equation} 
The last equations can be easily integrated when $\alpha$ and $\beta$ are constants. Under this assumption, we get that the potential $V(\phi)$ and the coupling function $f(\phi)$ are given by expressions of the form\footnote{In the case of varying  $\alpha$ and $\beta$, the autonomous set is not closed and we would be forced to introduce more dimensionless variables to close the system. Thus, we choose $\alpha$ and $\beta$ constant to keep our presentation simple.}
\begin{equation}
V(\phi) \propto 1/(\alpha  \phi)^{2}, \quad f(\phi) \propto (\alpha \phi)^{2\beta\phi/\left|\alpha\phi\right|},
\end{equation}
which correspond to common power law expressions typically used in the literature \cite{Copeland:2006wr,Yang:2012ht,Nozari:2013mba}.

\subsection{Analytical Fixed points}
\label{Section: AFP}

\begin{figure*}
\centering
\begin{minipage}[b]{.4\textwidth}
\includegraphics[width=\textwidth]{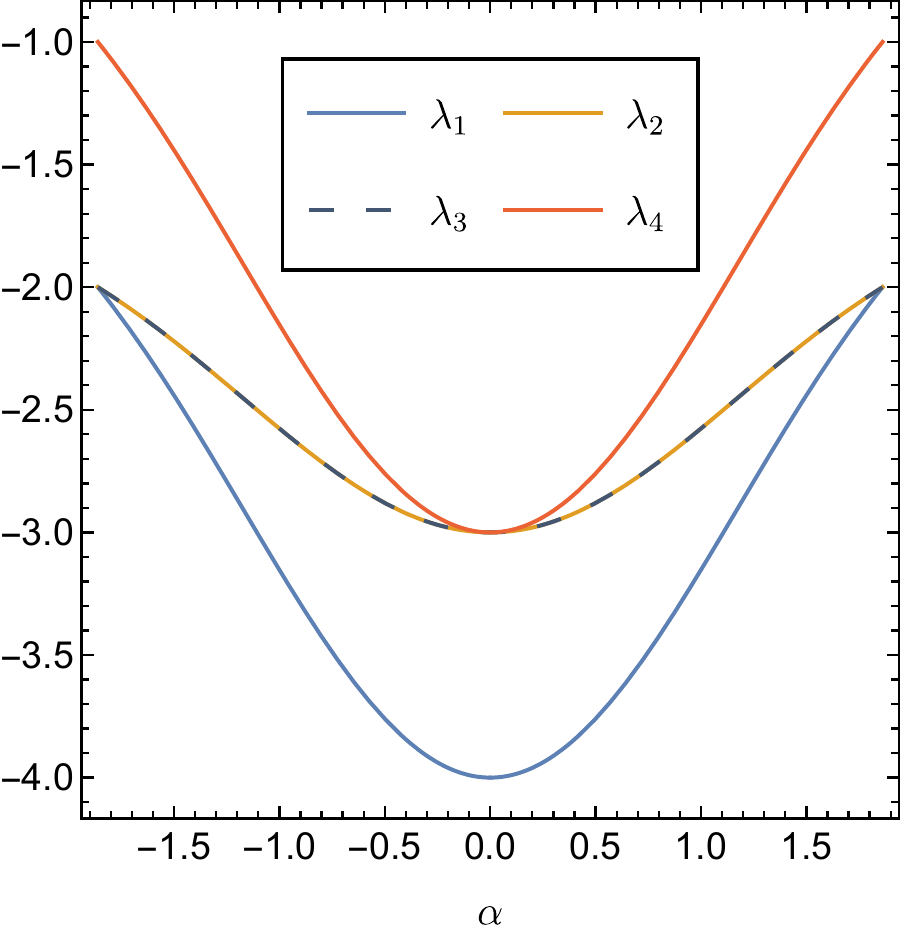}
\end{minipage} \hfill
\begin{minipage}[b]{.5\textwidth}
\includegraphics[width=\textwidth]{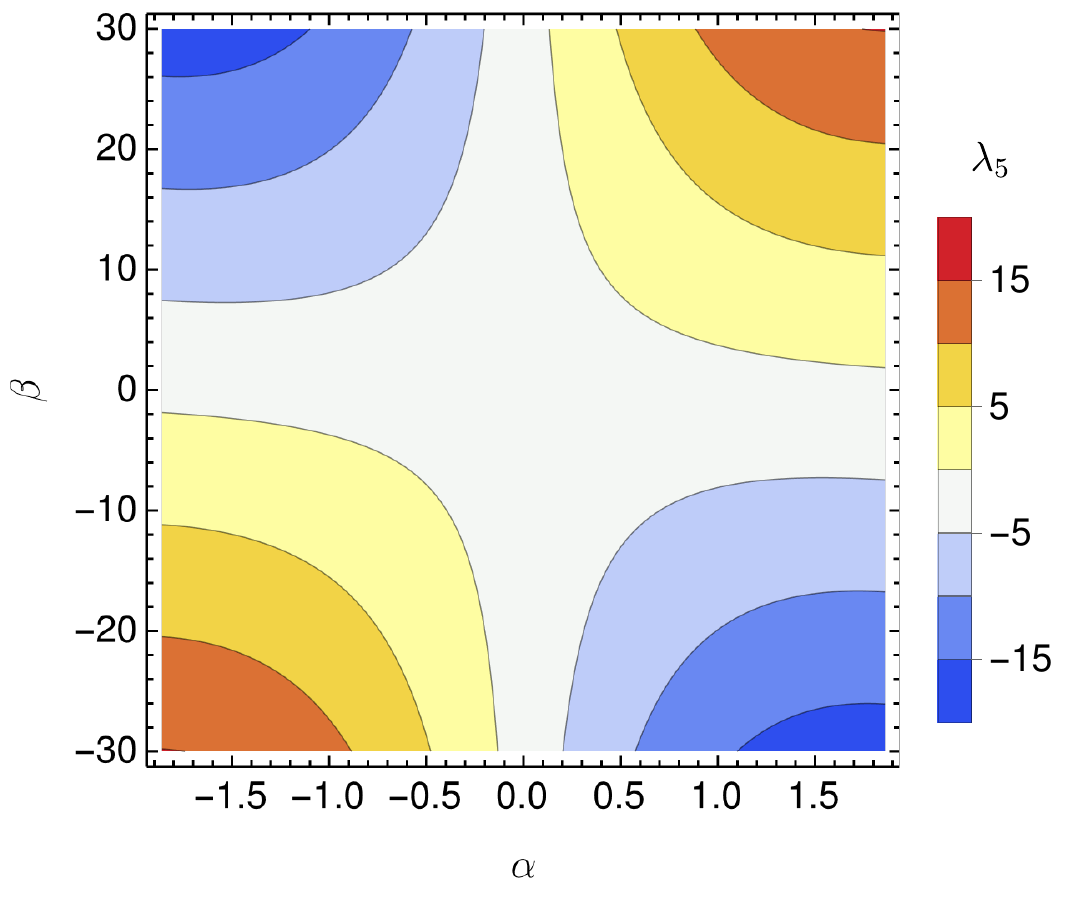}
\end{minipage}
\caption{Eigenvalues of the Jacobian matrix evaluated at (\emph{DE-I}). (Left) Note that $\lambda_{1, 2, 3, 4} < 0$ in the whole region where (\emph{DE-I}) is an accelerated solution, i.e., $w_{\text{eff}} < -1/3$. (Right) The eigenvalue $\lambda_5$ is negative in the regions given in Eqs. \eqref{Eq: DE-I Attractor 1} and \eqref{Eq: DE-I Attractor 2}.}
\label{Fig: Eigenvalues DEI}
\end{figure*}

The asymptotic behavior of the autonomous system in Eqs.~\eqref{Eq: x prime}-\eqref{Eq: Omega prime} is encoded in its fixed points, which can be found by setting $x' = 0$, $y' = 0$, $z' = 0$, $\Sigma' = 0$, and $\Omega'_r = 0$. This yields a set of algebraic equations given by
\begin{align}
0 &=\left( -1 + x^2 \right) \big\{ \alpha^2 y^4 + 2 \sqrt{3} \alpha  x y^3 + 3 x^2 y^2 \label{Eq: x prime polynomial} \\
& + \beta^2 x^2 z^4 - \beta^2 z^4 \big\}, \nonumber \\
0&=y\Big\{9 \left(x^2-1\right) y^4+\left(3 \Sigma ^2+\Omega_r+z^2+3\right)^2\label{Eq: y prime polynomial} \\
&+3 \alpha ^2 x^2 y^2+2 \sqrt{3} \alpha  x y \left(3 \Sigma ^2+\Omega_r+z^2+3\right)\Big\},\nonumber\\
0&=z\Big\{3 \beta ^2 x^2 y^2+\left[ \Sigma(3\Sigma-4) +\Omega_r+z^2-1\right]^2\label{Eq: z prime polynomial} \\
&-2 \sqrt{3} \beta  x y \left[\Sigma(3\Sigma-4)+\Omega_r+z^2-1\right]\nonumber\\
&+9 \left(x^2-1\right) y^4\Big\},\nonumber\\
0&=9 \Sigma ^6+\Sigma ^4 \left(6 \Omega_r+6 z^2-18\right)+24 \Sigma ^3 z^2\label{Eq: Sigma prime polynomial} \\
&+\Sigma ^2 \Big\{9 \left(x^2-1\right) y^4+\Omega_r^2-6 \Omega_r+z^4+2 \Omega_r z^2\nonumber\\
&-6 z^2+9\Big\}+16 z^4+\Sigma  \left(8 z^4+8 \Omega_r z^2-24 z^2\right),\nonumber\\
0&=\Omega_r \left\{9 \left(x^2-1\right) y^4+\left(3 \Sigma ^2+\Omega_r+z^2-1\right)^2\right\} \label{Eq: Omega_r prime polynomial},
\end{align}
where we have replaced the expression for $q$ in terms of the dynamical variables given in Eq.~\eqref{Eq: q deceleration parameter}, and some algebraic manipulations have been performed. Notice that these equations compose a polynomial system of degree greater than $4$, since Eq. \eqref{Eq: Sigma prime polynomial} is an equation of degree $6$ in the variable $\Sigma$. This means that the system is not analytically solvable in general due to the fundamental theorem of Galois theory \cite{ribes2010free}. Nonetheless, in the case $\Sigma = 0$, equation~\eqref{Eq: Sigma prime polynomial} implies $z = 0$, and Eq.~\eqref{Eq: z prime polynomial} is trivially satisfied. This is consistent with the fact that the vector field is the source of the anisotropy, as it can be seen in Eq.~\eqref{Eq: sigma}. The simplified Eqs.~\eqref{Eq: x prime polynomial}, \eqref{Eq: y prime polynomial} and \eqref{Eq: Omega_r prime polynomial} compose a system of degree reduced to 4, and thus analytical solutions exist.  

In order to do some analytical progress, in the following we neglect anisotropy, i.e., we consider $\Sigma = 0$ and $z = 0$, and study the fixed points of the simplified system in Eqs.~\eqref{Eq: x prime polynomial}, \eqref{Eq: y prime polynomial} and \eqref{Eq: Omega_r prime polynomial} relevant for the radiation era ($\Omega_r \simeq 1$, $ w_{\text{eff}} \simeq 1/3$), the matter era, ($\Omega_m\simeq 1$, $w_{\text{eff}} \simeq 0 $) and an isotropic DE era ($\Omega_{\text{DE}}\equiv\rho_{\text{DE}}/3M_{\text{Pl}}^2H^2\simeq 1$, ~$w_{\text{eff}}<-1/3$).

\begin{itemize}
\item (\emph{R}) Radiation dominance:
\end{itemize}
\begin{equation}
\label{Eq: R fixed point}
x = 0, \ y = 0, \ z = 0, \ \Sigma = 0, \ \Omega_r = 1,
\end{equation} 
with $\Omega_{\text{DE}} = 0$, $w_{\text{DE}}=-1$, $\Omega_m = 0$ and $w_{\text{eff}} = 1/3$. The eigenvalues of the Jacobian evaluated in this point are
\begin{equation}
2, \quad -1, \quad 1, \quad 0, \quad -3.
\end{equation}
Therefore, (\emph{R}) is a saddle. 

\begin{itemize}
\item (\emph{M}) Matter dominance:
\end{itemize}
\begin{equation}
x = 0, \ y = 0, \ z = 0, \ \Sigma = 0, \ \Omega_r = 0,\label{Eq: Matter dominance FP}
\end{equation}
with $\Omega_{\text{DE}} = 0$, $w_{\text{DE}}=-1$, $\Omega_m = 1$ and $w_{\text{eff}} = 0$. The eigenvalues of the Jacobian evaluated in this point are
\begin{equation}
-\frac{3}{2}, \quad \frac{3}{2}, \quad -1, \quad -\frac{1}{2}, \quad -3.
\end{equation}
Then, this point is a saddle. Note that the Jacobian has $4$ negative eigenvalues in this point, in contrast with the $2$ negative eigenvalues in the point (\emph{R}). This difference is crucial, since a correct expansion history requires a radiation dominated epoch followed by a matter dominated epoch. These eigenvalues reflect the fact that cosmological trajectories passing around the radiation dominated point (\emph{R}) can go to the ``more stable'' matter dominated point (\emph{M}).

\begin{itemize}
\item (\emph{DE-I}) Isotropic DE dominance:
\end{itemize}
\begin{equation*}
x = \mp \frac{\alpha \sqrt{\sqrt{36 + \alpha^{4}} -\alpha^{2}}}{3 \sqrt{2}}, \quad \Sigma = 0,
\end{equation*}
\begin{equation}
y = \pm \frac{\sqrt{\sqrt{36 + \alpha^{4}} - \alpha^{2}}}{\sqrt{6}}, \quad z = 0, \label{Eq: DE-I dominance FP}\\
\end{equation}
with $\Omega_{\text{DE}} = 1$, $\Omega_m = 0$, $\Omega_r = 0$ and 
\begin{equation}
w_{\text{DE}} = -1 + \frac{1}{18} \alpha^{2} \left(- \alpha^{2} + \sqrt{36 + \alpha^{4}}\right).
\end{equation}
Since $w_{\text{DE}} = w_\text{eff}$ in this point, the condition for accelerated expansion ($w_{\text{eff}} < -1/3$) is satisfied when 
\begin{equation}
|\alpha| < 3^{1 / 4} \sqrt{2} \approx 1.86121.
\end{equation}

In this case, the eigenvalues are given by large expressions which we present in Appendix \ref{App: DEI Eigenvalues}. From those expressions, we see that just one eigenvalue depends on $\alpha$ and $\beta$, while the remaining four eigenvalues depend only on $\alpha$. In the left panel of FIG.~\ref{Fig: Eigenvalues DEI}, we see that $\lambda_{1, 2, 3, 4} < 0$ in the region where (\emph{DE-I}) is an accelerated solution. In the right panel, we plot $\lambda_5$ in the same interval for $\alpha$, as in the left-panel, and we choose $\beta \in [-30, 30]$ as a representative region for this parameter. We find that $\lambda_5$ is negative when $\alpha = 0$ or in the regions
\begin{align}
0 < \alpha \quad &\land \quad \beta < -\frac{\alpha}{3} + \frac{2}{3} \sqrt{\frac{36 + \alpha^4}{\alpha^2}}, \label{Eq: DE-I Attractor 1} \\
\alpha < 0 \quad &\land \quad \beta > -\frac{\alpha}{3} - \frac{2}{3} \sqrt{\frac{36 + \alpha^4}{\alpha^2}}. \label{Eq: DE-I Attractor 2}
\end{align}
Therefore, (\emph{DE-I}) is an attractor in these regions. By looking at Eqs. \eqref{Eq: x prime}-\eqref{Eq: Omega prime}, we note that the autonomous system is invariant under the transformation $\{x \rightarrow -x, \alpha \rightarrow -\alpha, \beta \rightarrow -\beta \}$. This symmetry is reflected in FIG.~\ref{Fig: Eigenvalues DEI} and  in the regions in Eqs. \eqref{Eq: DE-I Attractor 1} and \eqref{Eq: DE-I Attractor 2}.

\subsection{Numerical Fixed points}

As shown in the last section, the set of algebraic equations in Eqs.~\eqref{Eq: x prime polynomial}-\eqref{Eq: Omega_r prime polynomial} does not have analytical solutions when $\Sigma \neq 0$. Therefore, our numerical setup can be useful for further analysis of the model. 

Despite early anisotropies being expected to be insignificant given the homogeneity of the CMB \cite{Planck:2019evm}, late-time anisotropies sourced by a dark energy component are not discarded by observations \cite{Campanelli:2010zx, Amirhashchi:2018nxl}. In the last section, we analytically found isotropic radiation, matter, and dark energy dominated points (\emph{R}), (\emph{M}), and (\emph{DE-I}). Then, our numerical search will be focused on anisotropic accelerated solutions. Moreover, it is possible to find initial conditions, in the deep radiation epoch, ensuring a proper radiation era followed by a standard matter era, given that the point (\emph{M}) is ``more stable'' than the point (\emph{R}). Hence, in the following, we will neglect the radiation component, i.e., $\Omega_r = 0$. This assumption will simplify our numerical treatment.

Now, we proceed with the implementation of the numerical setup explained in Sec.\ref{Sec: Dynamical systems: numerical approach}. Firstly, we have to choose a specific window parameter where the stochastic search will be performed. Having in mind that $|\alpha| < 1.86121$ for (\emph{DE-I}) being an accelerated solution, we choose $\alpha \in [-30,30]$ and $\beta \in [0,30]$; more precisely,
\begin{equation}
\{\alpha, \beta\} \in [-30,30]\times[0,30].
\label{Eq: search region}
\end{equation} 
Note that unlike $\alpha$, $\beta$ takes on nonnegative values since the autonomous system enjoys the symmetry $\{x \rightarrow -x, \alpha \rightarrow -\alpha, \beta \rightarrow -\beta \}$. Secondly, we impose the following physically motivated conditions
\begin{equation}
0 \leq \Omega_m \leq 1, \quad 0 \leq \Omega_\text{DE} \leq1.
\label{Eq: Physical Constraint}
\end{equation}
Therefore, any point in the parameter space is cataloged as a ``non-viable solution'' if all the found fixed points do not obey these physical constraints. If at least one of the corresponding fixed points meets all the conditions, the point is cataloged as a ``viable solution''. Now, we generate $N \sim 10^4$ random points in the region specified in Eq.~\eqref{Eq: search region}. For each of this points, the set of algebraic equations in Eqs.~\eqref{Eq: x prime polynomial}-\eqref{Eq: Sigma prime polynomial} is solved, considering $\Omega_r = 0$, using the \texttt{NSolve} command of \texttt{Mathematica}.\footnote{Note that our numerical scheme is not tied to the \texttt{Mathematica} software or to the \texttt{NSolve} command. This step, namely, the solution of the algebraic equations, can be done using any other coding language or algebraic system solver.} The code for solving $\sim 10^4$ algebraic systems takes approximately 45 minutes running on 136 parallelized mixed physical cores. This code is available on \href{https://github.com/sagaser/Numerical-dynamical-analysys}{GitHub}.\footnote{\href{https://github.com/sagaser/Numerical-dynamical-analysys}{https://github.com/sagaser/Numerical-dynamical-analysys}}

\begin{figure}[t!]
\centering
\begin{tikzpicture}
    \node at (0,0) {\includegraphics[width=0.48\textwidth]{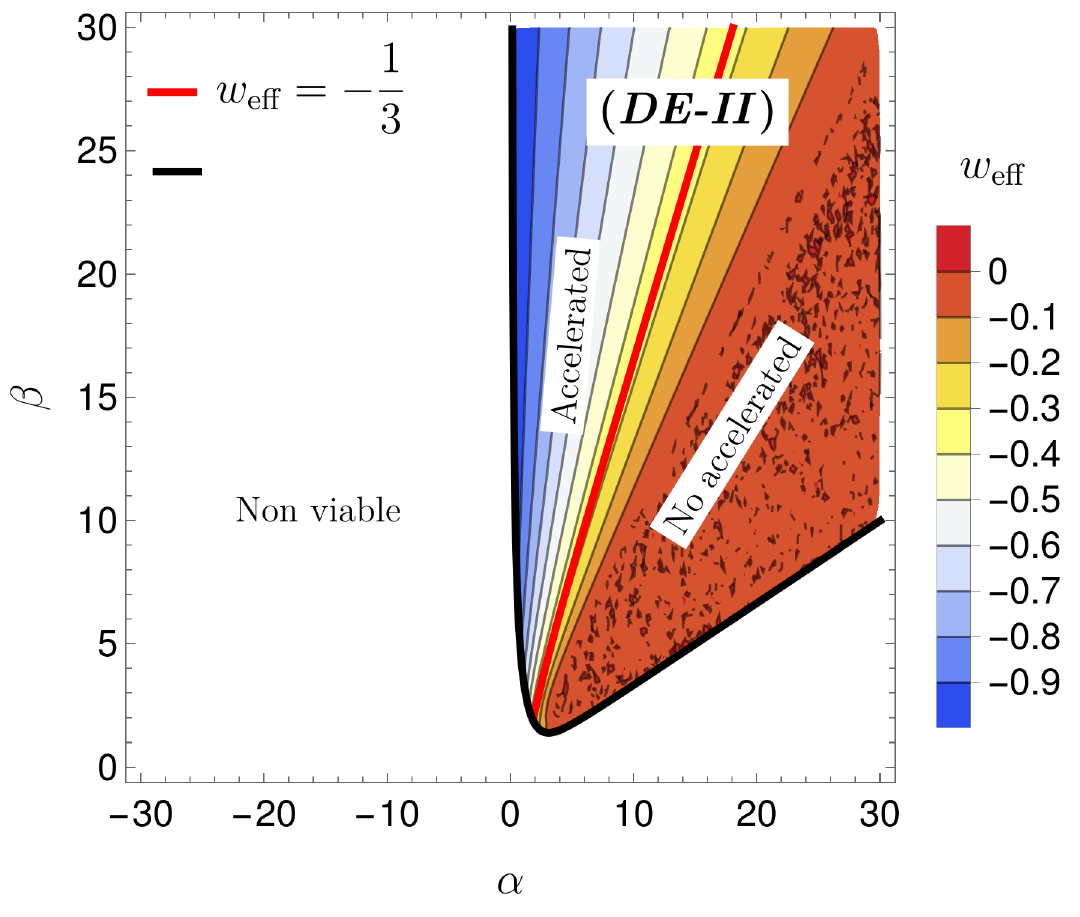}};
    \node at (-1.9,2.2) {Eq.~\eqref{Eq: Curve Viable Non-Viable}};
\end{tikzpicture}
\caption{Region in the parameter space where the stochastic search was performed. This region is divided in ``non-viable'' solutions and (\emph{DE-II}). Points in (\emph{DE-II}) match the physical constraints in Eq.~\eqref{Eq: Physical Constraint}. These regions are separated by the black curve whose expression is given in Eq.~\eqref{Eq: Curve Viable Non-Viable}. The red line represents the points for which $w_{\text{eff}} = -1/3$, which further divides (\emph{DE-II}) into two regions: accelerated solutions and non-accelerated solutions.}
\label{Fig: Region Accelerated Solutions}
\end{figure}

The result of code runs are depicted in FIG.~\ref{Fig: Region Accelerated Solutions}, which we explain in the following. If all the solutions for a given point in the region in Eq.~\eqref{Eq: search region} do not satisfy the physical conditions in Eq.~\eqref{Eq: Physical Constraint}, this point in the parameter space is allocated in the white region called ``non viable''. The points filling this region are cosmologically irrelevant. In turn, if at least one of the solutions for a given point matches all the physical constraints, this point is allocated in the ``viable region'' which is called (\emph{DE-II}). We found that these regions, non viable and (\emph{DE-II}), are separated by the curve
\begin{equation}
\beta = - \frac{\alpha}{3} + \frac{2}{3} \sqrt{\frac{36 + \alpha^4}{\alpha^2}},
\label{Eq: Curve Viable Non-Viable}
\end{equation}
which is the same curve dividing the regions where (\emph{DE-I}) is an accelerated solution [see Eqs. \eqref{Eq: DE-I Attractor 1} and \eqref{Eq: DE-I Attractor 2}]. Region (\emph{DE-II}) can be further divided into ``accelerated'' and ``non accelerated'' solutions by the line $w_{\text{eff}}(\alpha,\beta) = -1/3$. The color code in FIG.~\ref{Fig: Region Accelerated Solutions} shows that it is possible to get $w_\text{eff} \sim -1$ for some values of $\alpha$ and $\beta$. The existence of (\emph{DE-I}) and the ``accelerated'' region in (\emph{DE-II}) ensures an expanding universe at an accelerated rate, which can be isotropic or anisotropic.
  
\begin{figure}[t!]
\centering
\includegraphics[width=0.48\textwidth]{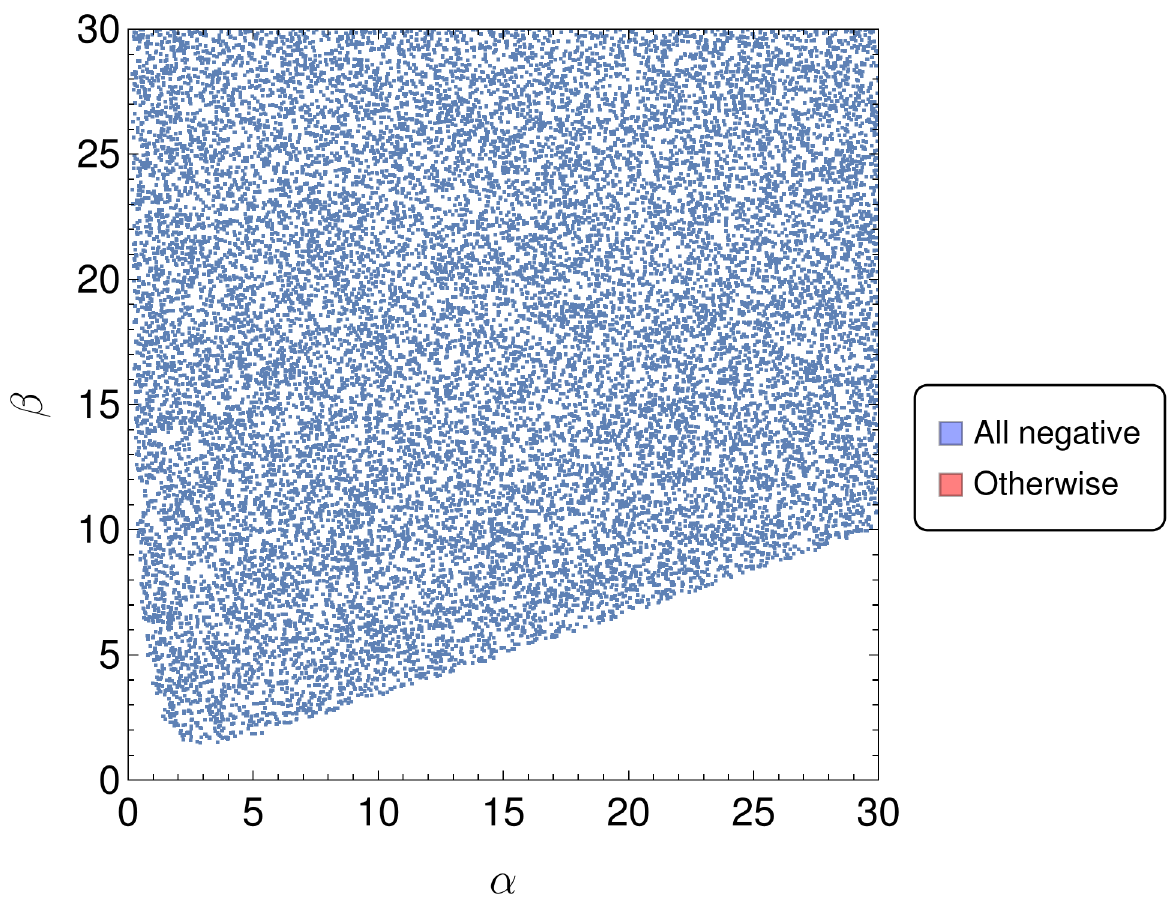}
\caption{Determination of the stability of the numerical fixed points conforming the region (\emph{DE-II}). A blue point represents an attractor point, which means that all the corresponding eigenvalues of at least one of the available Jacobian matrices are negative. Red points represent non-attractor points, namely, saddles or sources. Interestingly, there are no red points, meaning that (\emph{DE-II}) is an attractor in its own region of existence.} 
\label{Fig: Eigenvalues DE-II}
\end{figure}

\begin{figure}[t!]
\centering
\includegraphics[width = 0.9\linewidth]{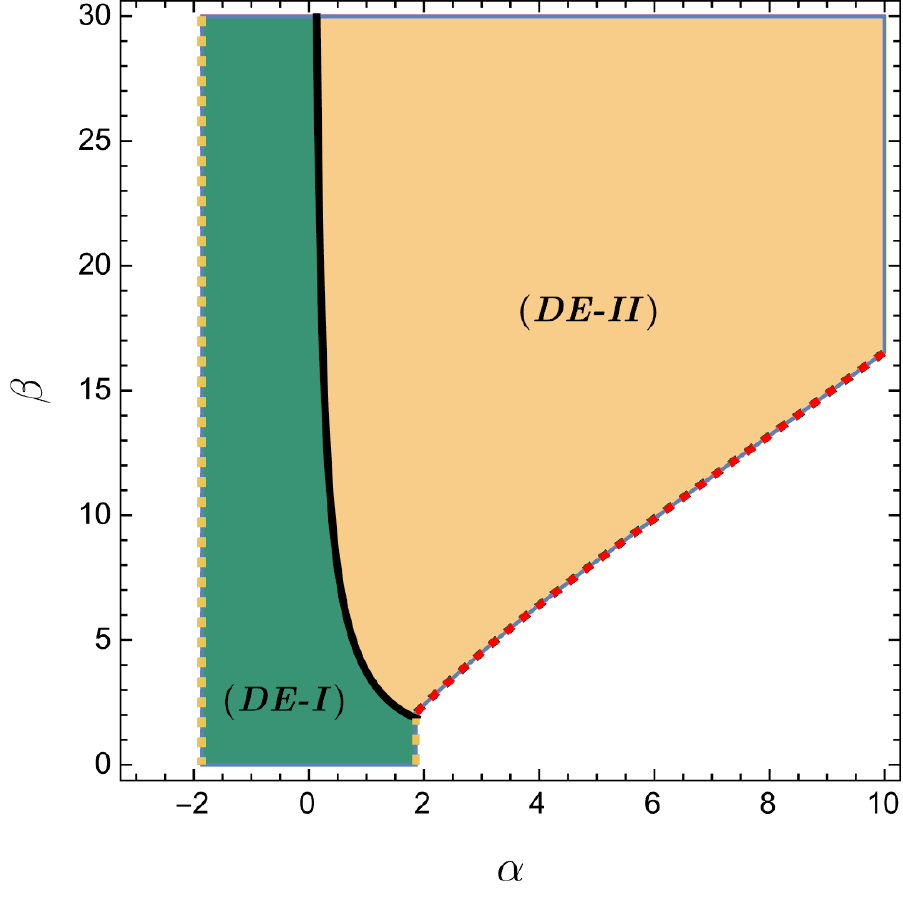}
\caption{Regions where (\emph{DE-I}) and (\emph{DE-II}) are attractors and accelerated solutions. The black line is the bifurcation curve given in Eq. \eqref{Eq: Curve Viable Non-Viable}.} 
\label{Fig: DEI and DEII}
\end{figure}

The stability of (\emph{DE-II}) can be determined following the step 4 in Sec.~\ref{Sec: Dynamical systems: numerical approach}. We compute all the available Jacobian matrices (and their eigenvalues) for each of the points in (\emph{DE-II}). We expect that at least a portion of the ``accelerated'' region in (\emph{DE-II}) could be an attractor of the system. As mentioned in Sec.~\ref{Sec: Dynamical systems: numerical approach}, a point in the parameter space yield attractor solutions if all the eigenvalues of at least one of the corresponding Jacobian matrices are negative. As shown in FIG.~\ref{Fig: Eigenvalues DE-II}, this attractor condition is obeyed by all the points in (\emph{DE-II}). Therefore, (\emph{DE-II}) is an attractor inside its own region of existence. We summarize the possible attractors of the system in FIG.~\ref{Fig: DEI and DEII}. In this figure, we find out that Eq.~\eqref{Eq: Curve Viable Non-Viable} is the bifurcation curve that separates both attractors.

\section{Numerical Integration of the Autonomous Set}
\label{Sec: Numerical Integration of the Autonomous Set}

In order to check our claims about the asymptotic behavior of the system, in this section we numerically solve the full autonomous system in Eqs.~\eqref{Eq: x prime}-\eqref{Eq: Omega prime} for specific values of $\alpha$ and $\beta$. The initial conditions are set at a very high redshift, $z_r = 6.57 \times 10^7$, ensuring that cosmological trajectories start in the deep radiation epoch. Moreover, we assume that possible anisotropies can be sourced only at late-times when the contribution of DE is significant to the energy budget, hence $\Sigma_i = 0$. For the remaining variables we choose
\begin{equation}
x_i = 10^{-25}, \quad z_i = 10^{-15}, \quad \Omega_{ri} = 0.99995. 
\label{Eq: Initial Conditions}
\end{equation}
The value for $y_i$ is constrained by Eq. \eqref{Eq: 2º Friedman}. For $0 \leq \Omega_{mi} \leq 1$, $y_i$ have to satisfy $|y_i| \leq 0.00707$, so we choose 
\begin{equation}
 y_i = 2.01 \times 10^{-14},
\end{equation}
such that $\Omega_{mi} = 5.0 \times 10^{-5}$. We would like to point out that the values of $x_i$, $y_i$, and $z_i$ are chosen so small to avoid possible large contributions of dark energy during the radiation dominated epoch.

\subsection{Isotropic Dark Energy Attractor}
\label{Sec: Isotropic Dark Energy Attractor}
Firstly, we choose $\alpha$ and $\beta$ such that (\emph{DE-I}) is the attractor of the system:
\begin{equation}
\alpha = 0.5, \quad \beta = 0.1,
\end{equation}

\begin{figure}[t!]
\centering
\includegraphics[width = 0.85\linewidth]{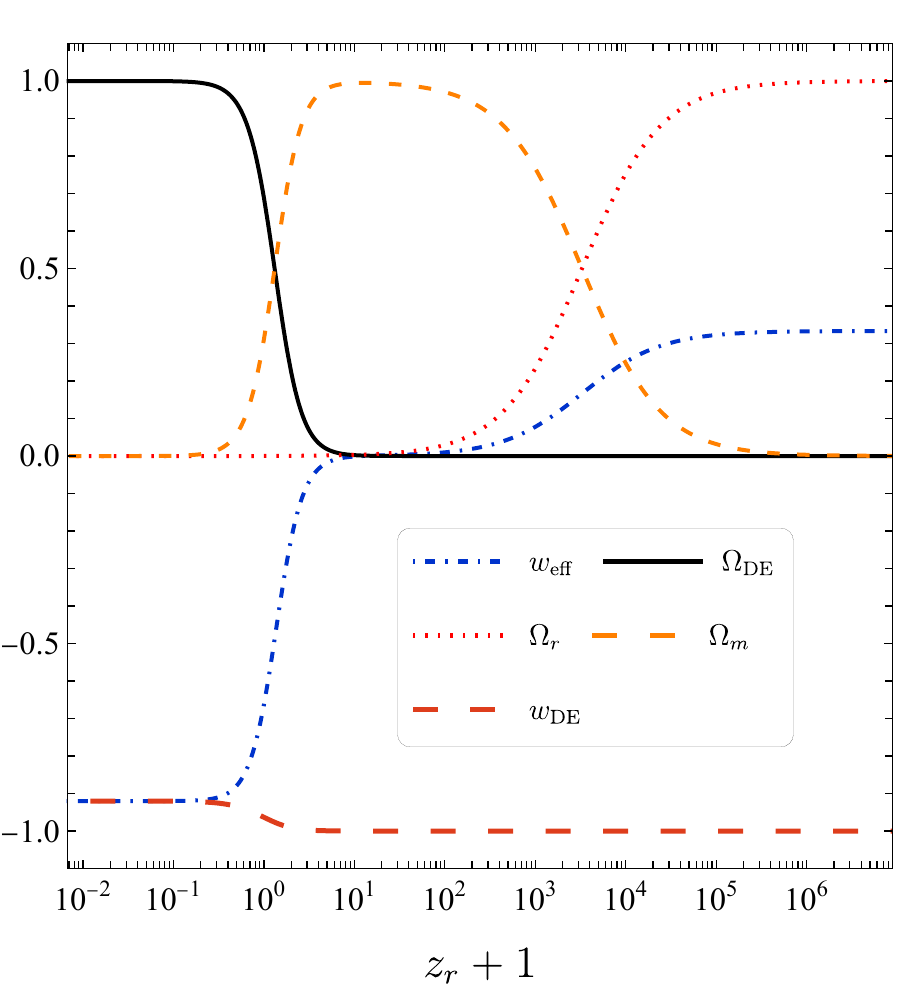}
\caption{Evolution of the density parameters, $w_\text{eff}$ and $w_\text{DE}$ during the whole expansion history. The initial conditions, Eq. \eqref{Eq: Initial Conditions}, were chosen in the deep radiation era at the redshift $z_r = 6.57 \times  10^7$. The universe passes through radiation dominance at early times (red dotted line), followed by a matter dominance (light brown dashed line), and ends in the DE dominance (black solid line) characterized by $w_{\text{eff}} = w_\text{DE} \approx -0.92$ (blue dot-dashed line and tangelo dashed line, respectively).} 
 \label{Fig: Isotropic Evolution}
\end{figure}

\begin{figure}[t!]
\centering
\includegraphics[width = \linewidth]{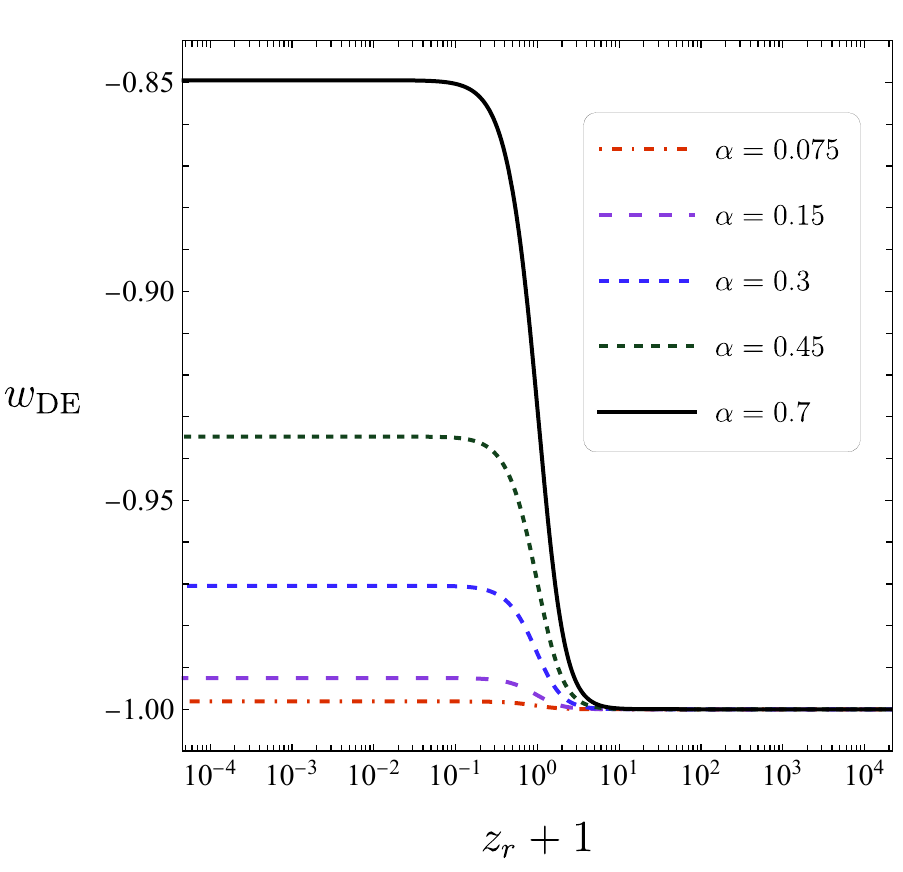}
 \caption{Time evolution of $w_{\text{DE}}$ for different values of the parameter $\alpha$, while $\beta = -90$. The initial conditions are the same given in Eq. \eqref{Eq: Initial Conditions}.} 
\label{Fig: DE-I Attractor}
\end{figure}

In FIG.~\ref{Fig: Isotropic Evolution}, we plot the evolution of the density parameters, the effective equation of state of the universe and the equation of state of DE as function of the redshift $z_r$. We can see that the early universe ($z_r>10^6$) is dominated by radiation (red dotted line). Then, at $z_r \approx 3200$, we have the radiation-matter transition, that is $\Omega_m \simeq \Omega_r$. After this transition we find $\Omega_{\text{DE}} \approx 9.89\times 10^{-10}$, which is in agreement with the BBN constraint $\Omega_{\text{DE}} < 0.045$ at $z_r = 1200$ \cite{Bean:2001wt}. From this transition to $z_r \approx 0.3$, the Universe is dominated by matter (light brown dashed line). The contribution of DE to the energy budget at $z_r = 50$ is $\Omega_{\text{DE}} = 1.74 \times 10^{-5}$, which agrees with the CMB constraint $\Omega_{\text{DE}}< 0.02$ at this redshift \cite{Planck:2018vyg}. This matter domination is followed by DE domination (black solid line). The expansion of the Universe speeds up since $w_\text{eff} < -1/3$ (blue dot-dashed line). Note  that the behavior of the equation of state of DE is in perfect agreement with what expected, i.e., it is equal to $-1$ during the radiation and matter epochs, and takes on a different value during the DE domination. This value depends only on $\alpha$ and in this case is given by $w_{\text{eff}} = w_{\text{DE}} \simeq -0.92$. In FIG.~\ref{Fig: DE-I Attractor}, we plot $w_\text{DE}$ for several values of $\alpha$ keeping $\beta$ fixed, where we can see that $w_\text{DE} \rightarrow -1$ when $\alpha \rightarrow 0$, as expected.

We want to point out that we verified that $w_\text{DE}$ does not depend on $\beta$ and also that $\Sigma = 0$ during the whole expansion history. This had to be so, since (\emph{DE-II}) does not exist when (\emph{DE-I}) is an attractor. However, the conversely is not true. In general, when (\emph{DE-II}) is the attractor of the system, (\emph{DE-I}) exist as a saddle, and the Universe could expand isotropically during a brief period of time.  

\subsection{Anisotropic Dark Energy Attractor}

For (\emph{DE-II}) to be the attractor of the system, we choose the following parameters
\begin{equation}
\alpha=0.5, \quad \beta=80,
\end{equation}
and we use the same initial conditions as in Eq. \eqref{Eq: Initial Conditions}.

\begin{figure}[t!]
\centering
\includegraphics[width = 0.85\linewidth]{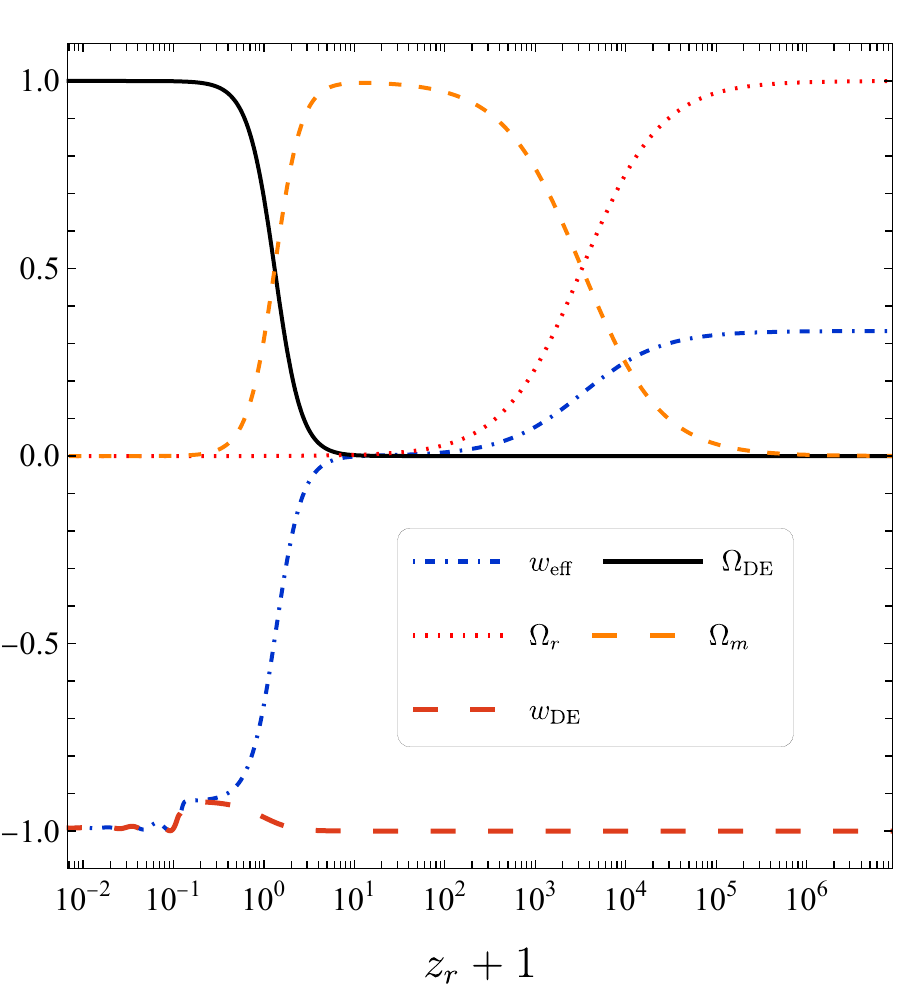}
\caption{Evolution of the density parameters, $w_\text{eff}$ and $w_\text{DE}$ during the whole expansion history for an anisotropic accelerated attractor. The initial conditions are the same as in FIG.~\ref{Fig: Isotropic Evolution}. The most appreciable difference with respect to the isotropic accelerated case (shown in FIG.~\ref{Fig: Isotropic Evolution}) is that $w_\text{DE} \approx -1$ oscillates at late times.} 
 \label{Fig: Anisotropic Evolution}
\end{figure}

In FIG.~\ref{Fig: Anisotropic Evolution}, we plot the evolution of the density parameters, the effective equation of state and the equation of state of the DE as function of the redshift $z_r$. The most prominent difference between this case and the (\emph{DE-I}) attractor case (shown in FIG.~\ref{Fig: Isotropic Evolution}) is the oscillatory behavior of $w_{\text{DE}}$ at late times. As mentioned in Ref. \cite{Orjuela-Quintana:2021zoe}, these oscillations occur when the kinetic term of the scalar field is comparable to the vector density, turning the equation of motion of the scalar field (Eq. \eqref{Eq: Eq phi} in this case) into an equation describing a damped harmonic oscillator. 

\begin{figure*}
\centering
\begin{minipage}[b]{.472\textwidth}
\includegraphics[width=\textwidth]{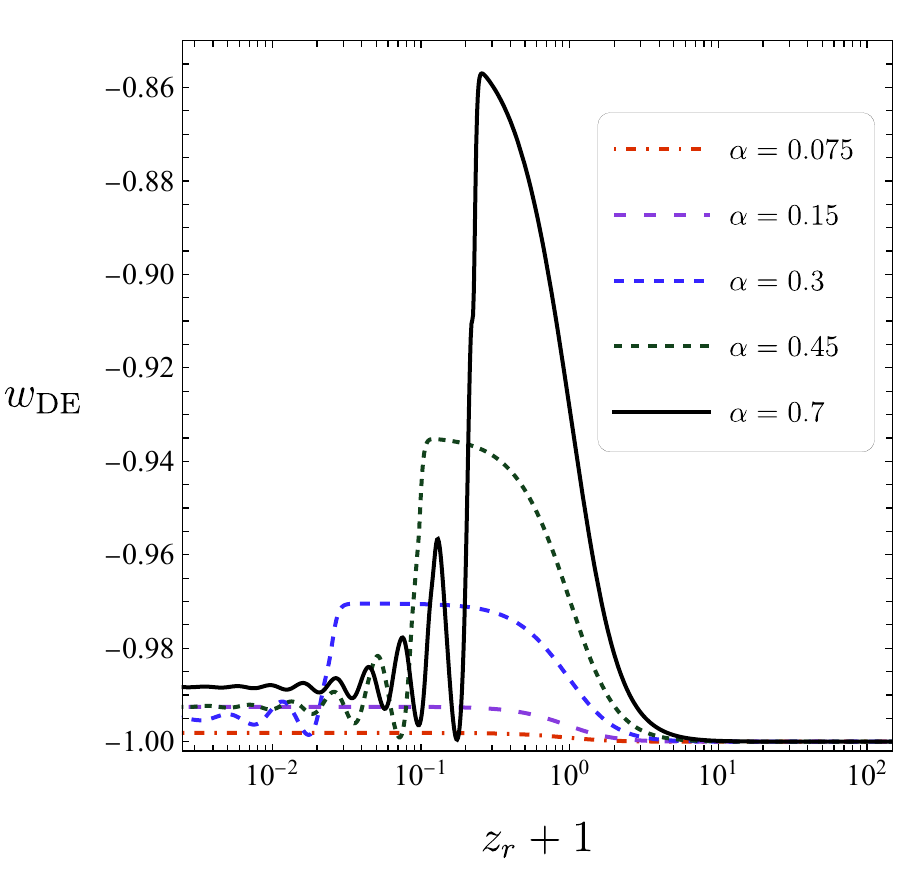}
\end{minipage} \hfill
\begin{minipage}[t]{.445\textwidth}
\includegraphics[width=\textwidth]{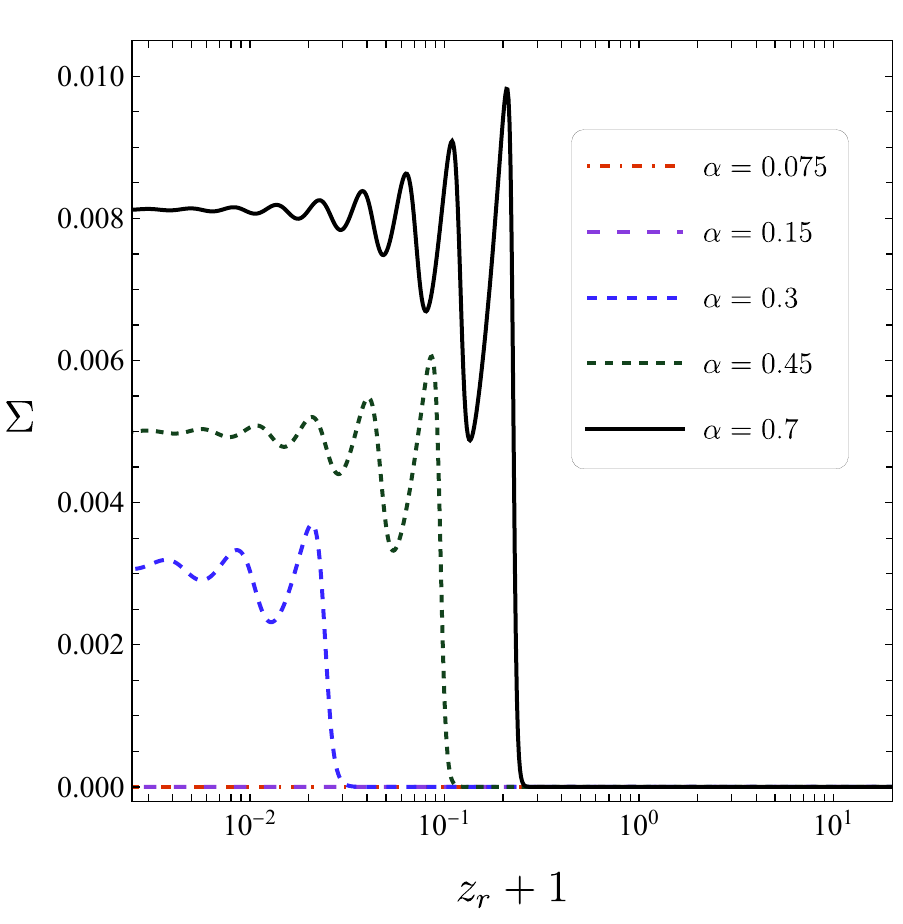}
\end{minipage}
\caption{Late-time evolution ($z_r < 100$) of the equations of state of DE $w_{\text{DE}}$ and the shear $\Sigma$ for different values of the parameter $\alpha$, while $\beta = 80$. The initial conditions are the same given in Eq. \eqref{Eq: Initial Conditions}. Note that the amplitude of the oscillations grows with $\alpha$. As $w_\text{DE}$ as $\Sigma$ behave as damped oscillators until they reach the asymptotical value predicted by the numerical fixed point (\emph{DE-II}).}
\label{Fig: Dark Energy Alpha}
\end{figure*}

\begin{figure*}
\centering
\begin{minipage}[b]{.472\textwidth}
\includegraphics[width=\textwidth]{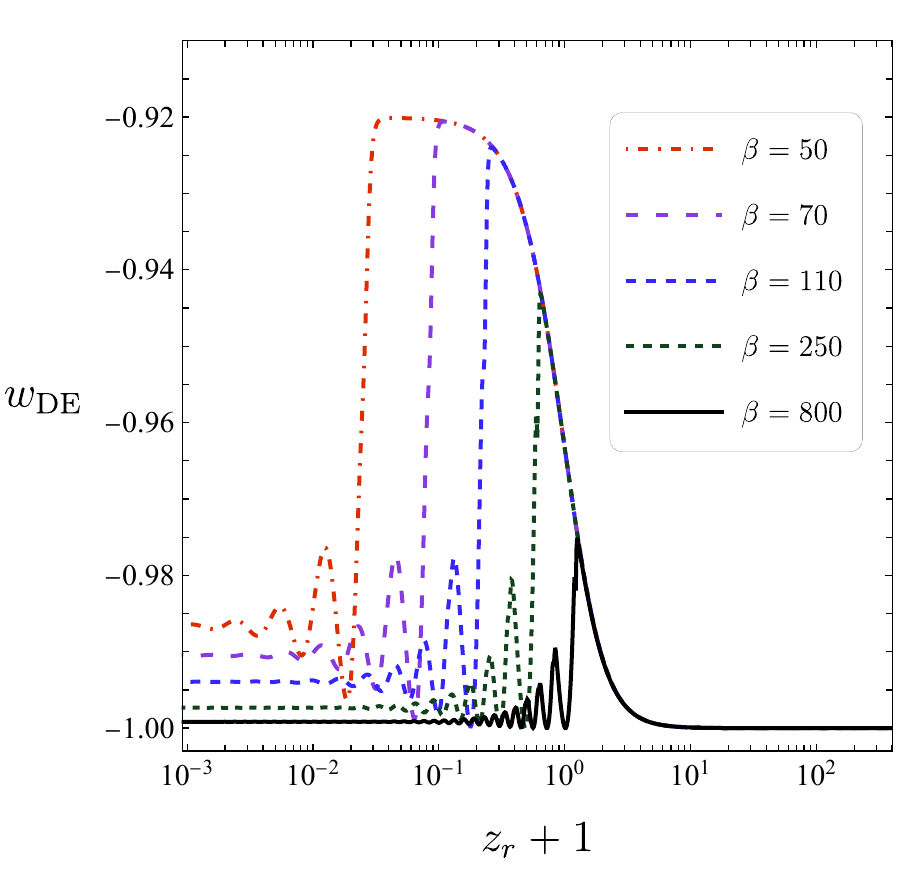}
\end{minipage} \hfill
\begin{minipage}[b]{.445\textwidth}
\includegraphics[width=\textwidth]{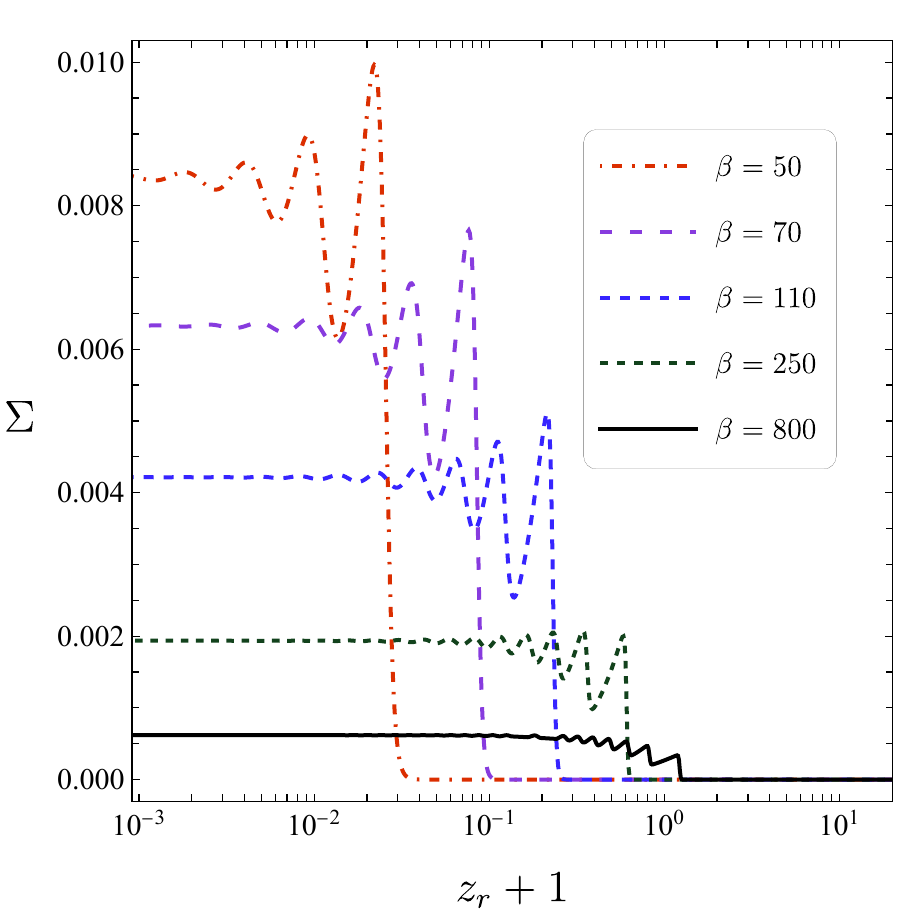}
\end{minipage}
\caption{Late-time evolution ($z_r < 100$) of the equations of state of DE (left) $w_{\text{DE}}$ and the shear $\Sigma$ (right) for different values of the parameter $\beta$, while $\alpha = 0.5$. The initial conditions are the same given in Eq. \eqref{Eq: Initial Conditions}. Note that the amplitude of the oscillations is greater for smaller $\beta$. As $w_\text{DE}$ as $\Sigma$ oscillate until they reach the asymptotical value predicted by the numerical fixed point (\emph{DE-II}).}
\label{Fig: Dark Energy Beta}
\end{figure*}

In FIG.~\ref{Fig: Dark Energy Alpha}, we plot the late-time evolution of $w_\text{DE}$ and $\Sigma$ for fixed $\beta = 80$ and varying the parameter $\alpha$. We observe that both $w_\text{DE}$ and $\Sigma$ oscillate until they stabilize to the value predicted in the numerical fixed point (\emph{DE-II}), as shown in Table \ref{tab: Predicted values}. Note that the amplitude of these oscillations grows with $\alpha$. We also investigate the behavior of $w_\text{DE}$ and $\Sigma$ when $\alpha$ is fixed and $\beta$ varies, as shown in FIG.~\ref{Fig: Dark Energy Beta}. In this case, the amplitude of the oscillations grows while $\beta$ decreases. Note in FIG.~\ref{Fig: Dark Energy Alpha} that when $w_\text{DE}$ deviates from $-1$ it grows until a value which mainly depends on $\alpha$. This indicates that the Universe is crossing the point (\emph{DE-I}) which is a saddle. When $\alpha$ is fixed, as in FIG.~\ref{Fig: Dark Energy Beta}, the time spent by the Universe crossing (\emph{DE-I}) depends on $\beta$. For larger values of $\beta$, the coupling between the tachyon field and the vector field is stronger, and thus oscillations start earlier. 

\begin{table}[h!]
    \centering
      \caption{Predicted values of $\Sigma$ and $w_{\text{DE}}$ as a function of each $\alpha$ and $\beta$ in the region where the numerical approach was applied and solved with a mean error of $20\%$ in $\Sigma$.}
\begin{tabular}{l|l|l|l}
\hline 
\hline
$\alpha$ & $\beta$ & $\Sigma\times 10^{-3}$ & $w_{\text {DE }}$ \\
\hline 
$0.5$ & $50$ & $5.58 \pm 2.70 $ & $-0.9867$ \\
$0.5$ & $70$ & $4.21 \pm 0.46 $ & $-0.9905$ \\
$0.5$ & $110$ & $2.81 \pm 0.56 $ & $-0.9939$ \\
$0.5$ & $250$ & $1.29 \pm 0.61 $ & $-0.9973$ \\
$0.5$ & $800$ & $41.2 \pm 25.1 $ & $-0.9992$ \\
\hline
$0.075$ & $80$ & $20.8 \pm 0.1 $ & $-0.9988$ \\
$0.15$ & $80$ & $83.3 \pm 0.1 $ & $-0.9975$ \\
$0.3$ & $80$ & $2.08 \pm 0.56 $ & $-0.9950$ \\
$0.45$ & $80$ & $3.33 \pm 0.46 $ & $-0.9925$ \\
$0.7$ & $80$ & $5.40 \pm 2.70 $ & $-0.9884$
\end{tabular}

    \label{tab: Predicted values}
\end{table}

In summary, we have shown that the following cosmological trajectories exist:
\begin{center}
(\emph{R}) \ $\rightarrow$ \ (\emph{M}) \ $\rightarrow$ \ (\emph{DE-I})/(\emph{DE-II}),
\end{center}
where (\emph{DE-II}) is a region in the parameter space with no analytical description. At this point, we would like to stress that the analysis presented in this section serves as a check for the asymptotic behavior predicted from the numerical fixed point (\emph{DE-II}). Hence, it establishes the consistency of the numerical scheme proposed in this work, and its usefulness when portions of the parameter space of a cosmological model is unreachable through analytical means.

\section{Conclusions} 
\label{Sec: Conclusions}

In this work, we have put forward a numerical method to explore the parameter space of a cosmological model when no analytical fixed points are available. We applied our method to a specific model of anisotropic dark energy based on the interaction between a scalar tachyon field and a vector field in a Bianchi I background. We have explicitly shown that the anisotropic attractor of the system has no analytical description, given that the degree of the algebraic system from which this point must be computed is greater than 4 [see Eqs.~\eqref{Eq: x prime polynomial}-\eqref{Eq: Sigma prime polynomial}]. However, when the anisotropy is neglected, the system is reduced to a system of degree 4 and thus analytical solutions exist, which we presented in Eqs.~\eqref{Eq: R fixed point}, \eqref{Eq: Matter dominance FP} and \eqref{Eq: DE-I dominance FP}. In particular, our method allowed us to find the parameter space of the model where anisotropic accelerated solutions exist as attractors of the system, which we plot in FIG.~\ref{Fig: DEI and DEII}. Then, we checked the consistency of the method by numerically solving the full autonomous system in Eqs.~\eqref{Eq: x prime}-\eqref{Eq: Omega prime} for a particular set of initial conditions. As a last remark, we would like to stress on the generality of our method, as explained in Sec. \ref{Sec: Dynamical systems: numerical approach}, i.e., in principle, it can be applied to any DE scenario.

\section*{Acknowledgements}

This work was supported by Patrimonio Aut\'onomo - Fondo Nacional de Financiamiento para la Ciencia, la Tecnolog\'ia y la Innovaci\'on Francisco Jos\'e de Caldas (MINCIENCIAS - COLOMBIA) Grant No. 110685269447 RC-80740-465-202, projects 69723 and 69553.

\newpage

\appendix

\section{(\emph{DE-I}) Eigenvalues}
\label{App: DEI Eigenvalues}

The eigenvalues for the isotropic DE, section \ref{Section: AFP} are given by
\begin{widetext}
\begin{align}
 \lambda_1 &= \frac{1}{12} \left(-\sqrt{2} \sqrt{\alpha ^4+36} \Gamma +\sqrt{2} \alpha ^2 \Gamma -12\right)\\
 \lambda_2 &= \frac{1}{24} \left(-\sqrt{2} \sqrt{\alpha ^4+36} \Gamma +\sqrt{2} \alpha ^2 \Gamma -36\right)\\
 \lambda_3 &= \frac{1}{48}\left(-36+3 \sqrt{2} \alpha^2 \Gamma-3 \sqrt{2} \sqrt{36+\alpha^4} \Gamma\right. \\
&\notag -2\left(66 \alpha^8+\alpha^6\left(-66 \sqrt{36+\alpha^4}+32 \sqrt{2} \Gamma\right)+\alpha^2\left(-900 \sqrt{36+\alpha^4}+738 \sqrt{2} \Gamma\right)\right. \\
& \notag\left.\left.-162\left(-36+\sqrt{2} \sqrt{36+\alpha^4} \Gamma\right)-8 \alpha^4\left(-261+4 \sqrt{2} \sqrt{36+\alpha^4} \Gamma\right)\right)^{\frac{1}{2}}\right)\\
 \lambda_4 &= \frac{1}{48}\left(-36+3 \sqrt{2} \alpha^2 \Gamma-3 \sqrt{2} \sqrt{36+\alpha^4} \Gamma\right. \\
&\notag +2\left(66 \alpha^8+\alpha^6\left(-66 \sqrt{36+\alpha^4}+32 \sqrt{2} \Gamma\right)+\alpha^2\left(-900 \sqrt{36+\alpha^4}+738 \sqrt{2} \Gamma\right)\right. \\
& \notag \left.\left.-162\left(-36+\sqrt{2} \sqrt{36+\alpha^4} \Gamma\right)-8 \alpha^4\left(-261+4 \sqrt{2} \sqrt{36+\alpha^4} \Gamma\right)\right)^{\frac{1}{2}}\right)\\
 \lambda_5 &= \frac{1}{24} \left(2 \sqrt{\alpha ^4+36} \alpha  \beta -\sqrt{2} \sqrt{\alpha ^4+36} \Gamma -2 \alpha ^3 \beta +\sqrt{2} \alpha ^2 \Gamma -12\right)
\end{align}
\end{widetext}    
where $\Gamma\equiv\sqrt{\alpha ^4-\sqrt{\alpha ^4+36} \alpha ^2+18}$, $\lambda_1, \ldots, \lambda_4$ are all negative if $\alpha$ is a real number, and $\lambda_5$ is negative when Eq. \eqref{Eq: DE-I Attractor 1} or Eq. \eqref{Eq: DE-I Attractor 2} are satisfied.

\bibliography{Bibli.bib}
\end{document}